\documentclass[aps,prb,preprint,superscriptaddress]{revtex4-1}

\usepackage{graphicx}
\usepackage{amsmath,amssymb,mathrsfs}
\usepackage{hyperref}

\DeclareMathOperator{\sech}{sech}

\begin{document}

\title{Acoustic damping and dispersion in vitreous germanium oxide}

\author{Guillaume Guimbreti\`ere}
\affiliation{Universit\'e Montpellier 2, Laboratoire Charles Coulomb UMR 5221, F-34095, Montpellier, France}
\affiliation{CNRS, Laboratoire Charles Coulomb UMR 5221, F-34095, Montpellier, France}
\affiliation{CNRS-UPR 3079 CEMHTI, F-45071 Orl\'eans, France}

\author{Beno\^it Ruffl\'e}
\author{Ren\'e Vacher}
\affiliation{Universit\'e Montpellier 2, Laboratoire Charles Coulomb UMR 5221, F-34095, Montpellier, France}
\affiliation{CNRS, Laboratoire Charles Coulomb UMR 5221, F-34095, Montpellier, France}

\date{\today}

\begin{abstract}
New Brillouin scattering measurements of velocity and attenuation of sound in the hypersonic regime are presented. The data are analyzed together with the literature results at sonic and ultrasonic frequencies. As usual, thermally activated relaxation of structural entities describes the attenuation at sonic and ultrasonic frequencies. As already shown in vitreous silica, we conclude that the damping by network viscosity, resulting from relaxation of thermal phonons, must be taken into account to describe the attenuation at hypersonic frequencies. In addition, the bare velocity obtained by subtracting to the experimental data the effect of the two above mechanisms is constant for temperatures below 250~K, but increases almost linearly above, up to the glass transition temperature. This might indicate the presence of a progressive local polyamorphic transition, as already suggested for vitreous silica.
\end{abstract}
\maketitle

\section{Introduction}

In insulating solids, high frequency atomic vibrations are responsible for heat transport and thus determine the thermal properties. While the mechanisms leading to phonon damping in crystals are rather well understood, those responsible for sound attenuation in non-crystalline solids are still a matter of large interest. The main problem is that, due to the lack of periodicity, the exact nature of local and of collective vibrations is unknown.

At very low temperatures, it has been demonstrated that the dominant damping mechanism is a coupling of acoustic vibrations with tunneling units~\cite{ANDE1972,PHIL1972}. While the microscopic nature of those systems is not known, the phenomenological model involving resonance and relaxation of two-level tunneling systems (TLS) explains the temperature ($T$) and frequency ($\Omega$) dependence of sound velocity $v$ and attenuation $\alpha$~\cite{JACK1972,JACK1976}. In a large $T$ domain above 10~K the sound attenuation is dominated by a large broad peak which shifts to high $T$ with increasing frequency. This peak has been assigned to the coupling of acoustic vibrations with structural entities described as defects moving in asymmetric double-well potentials by thermally activated relaxation (TAR)~\cite{ANDE1955,HUNK1976,GILR1981}. This explains the $T$-dependence of $\alpha$ observed in most sonic and ultrasonic experiments. However, the $T$-dependence of the sound velocity in glasses such as As$_2$S$_3$ and As$_2$Se$_3$~\cite{CLAY1978} cannot be explained by this model: the behavior observed there is very similar to the decrease of sound velocity in crystals, where it is assigned to relaxation of thermal phonons by anharmonic interactions~\cite{AKHI1939,MARI1971}. This suggests that, even if the nature of thermal modes in glasses is unknown, and certainly different from propagating plane-wave phonons, a similar anharmonic mechanism can be present~\cite{VACH1981,FABI1999}. In fact, we have observed that Brillouin scattering measurements of $\alpha$ in vitreous silica at frequencies in the hypersonic range cannot be described satisfactorily by the TAR mechanism alone~\cite{VACH2005}. The excess to the TAR contribution was assigned to anharmonicity. The measurements were extended to frequencies up to 400~GHz~\cite{DEVO2008,*RUFF2011,*AYRI2011,*KLIE2011}. The results are well described by the sum of the contributions of the three above mechanisms, TLS, TAR and anharmonicity. Due to the different $T$ and $\Omega$ dependence of those mechanisms, anharmonicity tends to dominate above 100~GHz. In vitreous silica, $v$-SiO$_2$, we have performed a simultaneous fit of measurements of $\alpha$ and $v$ at low $T$ in a large frequency domain, which allowed to extract the parameters of the models. Calculating the velocity dispersion up to high $T$ and subtracting the contributions given by TAR and anharmonicity from the experimental results, one expects to get a constant bare velocity. In fact, we observed that the bare velocity shows an almost linear increase above 100~K. This was explained by a temperature dependent microscopic structural change leading to $T$-hardening~\cite{POLI2002,HUAN2004,AYRI2011a}.

A well-known characteristic of glasses is the plateau observed in the thermal conductivity in the 5-10~K range~\cite{ZELL1971}. This plateau suggests that a strong damping mechanism limits the mean free path of the acoustic waves responsible for thermal transfer~\cite{GRAE1986}. In this temperature range, the frequency of dominating thermal acoustic waves is around 1~THz~\cite{RAYC1989}. A mechanism leading to this strong damping must be invoked in addition to those discussed above. Using Brillouin scattering of X-rays to investigate acoustic waves in the THz range, we have given evidence for an attenuation increasing as $\Omega^4$ in two glasses leading to a Ioffe-Regel crossover~\cite{RUFF2003,*RUFF2006}. Above the frequency of this crossover, plane acoustic waves can no longer propagate~\cite{GRAE1986}.

The association of the mechanisms discussed above gives a satisfactory description of most of the $\alpha$ and $v$ measurements in vitreous silica. The purpose of the present paper is to extend this description to vitreous germanium oxide ($v$-GeO$_2$), another tetrahedrally coordinated glass.

\section{Material and Methods}

The samples used were kindly provided by Dr. J.~C. Lasjaunias, from the {\it Centre de Recherche sur les Tr\`es Basses Temp\'eratures} in Grenoble, France. Their optical transparency and homogeneity are excellent. A rough estimate by infrared absorption at 3560~cm$^{-1}$ gives a value of $\simeq$~100~ppm for the OH$^-$ content. We have observed the influence of the thermal history on $\alpha$ measurements in the hypersonic range: differences of about 20\% were observed between fast quenched and annealed samples. The latter have been used for this study. The velocity and attenuation of acoustic waves in the hypersonic frequency range were measured by Brillouin scattering~(BS). The spectrometer used for the experiments has been described elsewhere~\cite{SUSS1979,*VACH2006}. Its resolving power of about 2$\times$10$^7$ is large enough to allow precise measurements of Brillouin frequency shifts and linewidths. The 514.5~nm line of a single-frequency argon-ion laser was used. The measurements were performed in the backscattering geometry were only longitudinal modes are active in BS for isotropic media.

\begin{figure}
\includegraphics[width=8.5cm]{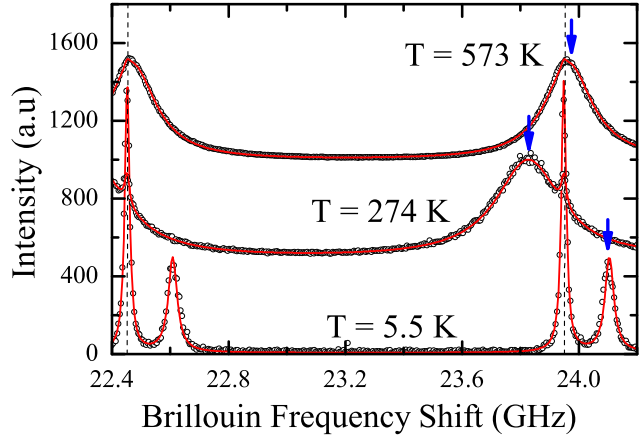}
\caption{Typical spectra registered at three temperatures. A full interference order of the confocal FP analyzer is shown. The Stokes Brillouin peaks at interference order $p$ are indicated by the arrows whereas dashed lines indicate the elastic peaks at orders $p$+15 and $p$+16. The red lines result from fits explained in the text.}
\label{Fig1}
\end{figure}

In order to reach the high resolving power needed with a sufficient luminosity, we use a confocal Fabry-Perot~(FP) interferometer with a free spectral range of 1.497~GHz. Some typical spectra are shown in Fig.~\ref{Fig1}. The Brillouin frequency shift is about 24~GHz, which means that the elastic and the Brillouin lines are observed at different interference orders. The frequency-shift scale on the figure is that for the Brillouin line and the arrows indicate the central frequency of the latter. The excellent signal-to-noise ratio observed for the spectra is reached in about 20 minutes at 5.5~K whereas 5 minutes are sufficient at room-$T$. From this figure, it is clear that both the frequency shift and the linewidth can be measured with a high accuracy in the whole temperature range. To extract these parameters, the Brillouin spectra are fitted to a damped harmonic oscillator convoluted with the instrumental profile taken from the elastic line. The small parasitic broadening due to the aperture of collection ($\simeq 40$~mrad) is taken into account.

\section{Results and Data Analysis}

\begin{figure}
\includegraphics[width=8.5cm]{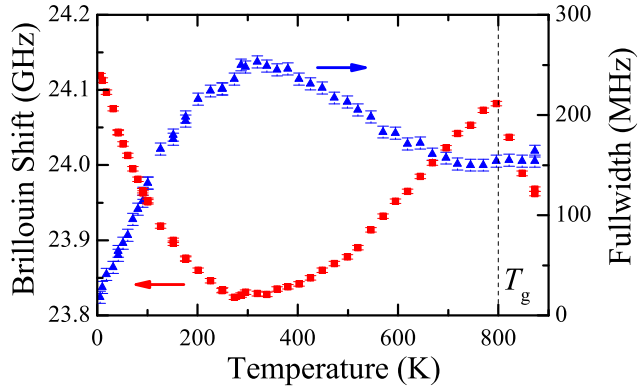}
\caption{Brillouin frequency shift $\delta\Omega_B/2\pi$ (squares, left scale) and fullwidth $\Gamma/2\pi$ (triangles, right scale) as a function of temperature. The vertical dashed line marks the glass transition temperature $T_g$.}
\label{Fig2}
\end{figure}

The results for $T$ from 5.5~K to 900~K are shown in Fig.~\ref{Fig2}. From the lowest $T$, the frequency shift $\delta\Omega_B/2\pi$ decreases strongly to a minimum around 300~K and increases above, up to the glass transition temperature $T_g\simeq$~800~K. At the latter $T$, a cusp is observed, as usual in glasses. The fullwidth at half maximum of the Brillouin line, $\Gamma/2\pi$, which is very small at low $T$, increases rapidly with increasing $T$ up to a maximum around 300~K, decreases above up to $\simeq$~700~K, and finally shows a plateau at large $T$, without noticeable change at $T_g$.

From the values of $\delta\Omega_B$, the velocity $v$ can be calculated using
\begin{equation}
\frac{\delta\Omega_B}{2\pi} = \frac{2nv}{\lambda_0} \sin\theta/2,
\end{equation}
where $n$ is the refractive index, $\lambda_0$ the wavelength of the incident light in vacuum and $\theta$ the scattering angle. The values of $n$ = 1.610 at 300~K and its $T$-dependence were taken from the literature~\cite{MAZU1985}. The attenuation $\alpha$ is obtained from $\Gamma$ measurements, $\alpha = \Gamma/v$. For comparison of attenuation measurements in a large $\Omega$ domain it is more convenient to use the internal friction:
\begin{equation}
Q^{-1} = \frac{\Gamma}{\delta\Omega_B}.
\end{equation}

A previous Brillouin scattering measurement of $v$ and $Q^{-1}$ in $v$-GeO$_2$ was performed earlier by \citet{HERT1998} in the $T$-range from 50~K to 750~K. Our results which are much more accurate agree with these measurements.

\section{Discussion}

\begin{figure}
\includegraphics[width=8.5cm]{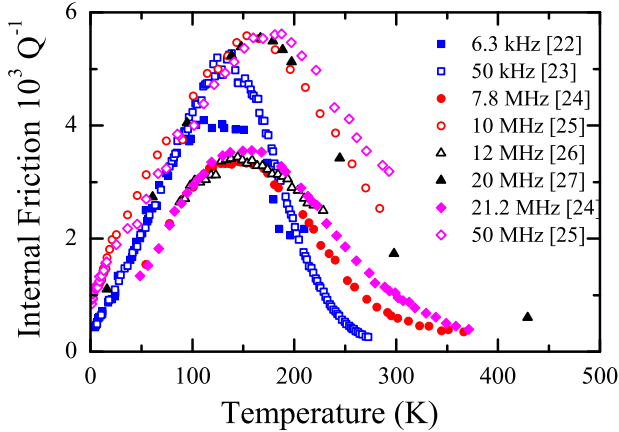}
\caption{Internal friction of $v$-GeO$_2$ as a function of temperature at several frequencies: 6.3~kHz~\cite{RAU1995}, 50~kHz~\cite{KOSU1996}, 7.8~MHz~\cite{STRA1964a}, 10~MHz~\cite{CARI2008b},~12 MHz~\cite{SAKA1989}, 20~MHz~\cite{KRAU1968}, 21.2~MHz \cite{STRA1964a}, and 50~MHz~\cite{CARI2008b}.}
\label{Fig3}
\end{figure}

In order to elucidate the contribution of the different damping mechanisms in vitreous germania, it is necessary to analyze simultaneously data measured in large $T$ and $\Omega$ domains. In comparison to the situation in vitreous silica, the data in $v$-GeO$_2$ are scarce. The results of sonic and ultrasonic measurements collected from the literature are shown in Fig.~\ref{Fig3}. As in most other glasses, a strong broad peak is observed in the $Q^{-1}$ measurements over the $T$, $\Omega$ domain shown. This indicates that the TAR mechanism is the main origin of the sound damping there. From our study of vitreous silica and also from preliminary calculations of the possible anharmonic damping in $v$-GeO$_2$, we know that this contribution is negligible in this $T$, $\Omega$ domain. This is also the case for $Q^{-1}$ from the lower $T$ up to 50~K in Brillouin scattering experiments. The temperature shift of the maximum with increasing $\Omega$ observed in Fig.~\ref{Fig3} allows to evaluate an average activation energy $V_0 \simeq$ 2400~K, as already shown by \citet{HERT1998}. It is more difficult to compare the amplitude of the curves, as measurements at almost the same $\Omega$ give appreciably different values. In particular, the measurements at 7.8, 12 and 21.2~MHz have almost the same amplitude, but are lower by a factor of about 1.7 than those at 10, 20 and 50~MHz. The comparison of amplitude from measurements performed on glass samples of different origin is generally difficult, as the thermal history and OH$^-$ content have an appreciable influence on the results. On the other hand, absolute calibration of $Q^{-1}$ measurements in ultrasonic experiments is always a difficult task. It is thus impossible to fix in the fit the same amplitude coefficient for the TAR contribution from the experiments shown in Fig.~\ref{Fig3}.

\begin{figure}
\includegraphics[width=8.5cm]{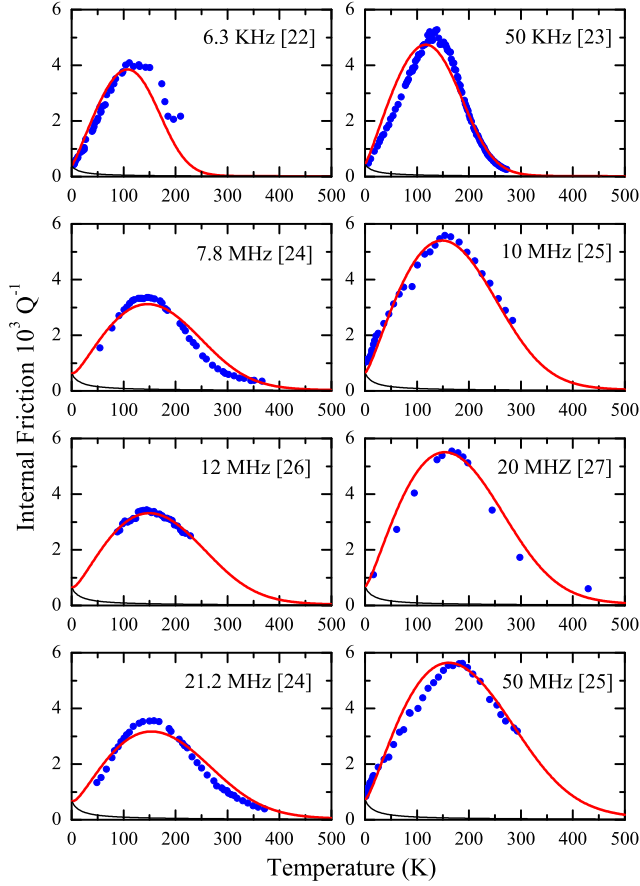}
\caption{Internal friction of $v$-GeO$_2$ in the sonic and ultrasonic frequency ranges fitted to the sum of TAR and TLS contributions as discussed in the text. Thin black lines show the small TLS contributions extracted from Refs. \onlinecite{RAU1995,CARI2008b}.}
\label{Fig4}
\end{figure}

The thermally activated defects are described by a distribution $P(\Delta,V)$ of asymmetric double-well potentials with a barrier height $V$ and an asymmetry $\Delta$. The contribution of the TAR mechanism to $Q^{-1}$ (see {\it e.g.} Ref. \onlinecite{PHIL1987}), $Q_{\rm TAR}^{-1}$, is given by
\begin{equation}
Q_{\rm TAR}^{-1} = \frac{\gamma^2}{\rho v^2 T} \int_{-\infty}^{\infty} d\Delta \int_{0}^{\infty} dV P(\Delta,V) \sech^2\frac{\Delta}{2 T} \frac{\Omega \tau}{1 + \Omega^2 \tau^2},
\label{Eq3}
\end{equation}
where $\gamma$ is a deformation potential, and $\rho$ is the mass density. In this equation, $\tau$ is the relaxation time for hopping within the double well. It is given by
\begin{equation}
\tau = \tau_0 \exp\frac{V}{T} \sech\frac{\Delta}{2 T},
\end{equation}
where $\tau_0$ is the inverse of an attempt frequency~\cite{TIEL1992}.

We have made a common fit of all $Q^{-1}$ measurements shown in Fig.~\ref{Fig3}, together with our Brillouin data at $T$ below 50~K. As a consequence of the above discussion on the amplitudes, we have fixed a common amplitude coefficient for measurements at 10, 20, 50~MHz and 24~GHz. This coefficient was left free for measurements at the other frequencies.

\begin{figure}
\includegraphics[width=8.5cm]{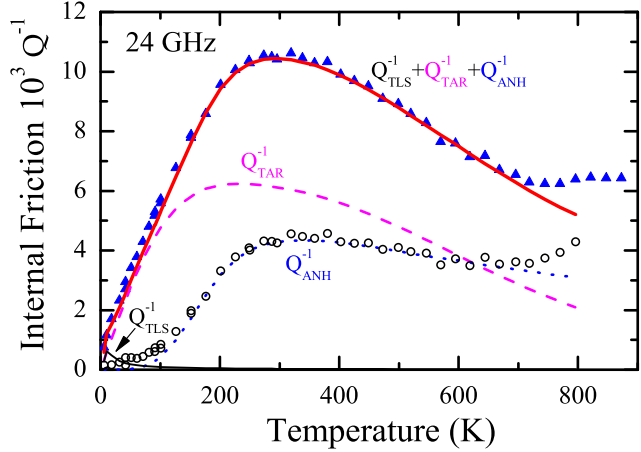}
\caption{(triangles) Internal friction of $v$-GeO$_2$ at hypersonic frequencies derived from the Brillouin linewidths. The dashed line shows the TAR contribution calculated from Eq.~(\ref{Eq3}) whereas the thin solid line is the TLS contribution. The circles represent the internal friction observed in Brillouin scattering after subtraction of $Q_{\rm TAR}^{-1}$ and $Q_{\rm TLS}^{-1}$. The dotted line is the calculated anharmonic contribution $Q_{\rm ANH}^{-1}$ as explained in the text. The solid line is the sum of these three contributions.}
\label{Fig5}
\end{figure}

For the distribution $P(\Delta,V)$, our first trial was a gaussian distribution for both parameters as in Ref.~\onlinecite{VACH2005}. The result of the fits disagrees with the experiments, giving a profile which is much broader than that observed. A better description was obtained by taking a sharper cut-off at $V_0$, {\it i.e.} an analytic form proportional to $\exp(-(V^2/2V_0^2)^2)$ for the distribution of barriers. Furthermore, the $Q^{-1}$ calculation at the frequencies of the experiments shown in Fig.~\ref{Fig3} is hardly sensitive to the cut-off $\Delta_0$ in the distribution of asymmetries as the corresponding relaxation times are too short to influence the acoustic damping. In contrast, the velocity change $\delta v$, given by 
\begin{equation}
\frac{\delta v}{v} = -\frac{1}{2} \frac{\gamma^2}{\rho v^2 T} \int_{-\infty}^{\infty} d\Delta \int_{0}^{\infty} dV P(\Delta,V) \sech^2\frac{\Delta}{2 T} \frac{1}{1 + \Omega^2 \tau^2},
\label{Eq5}
\end{equation}
is sensitive to the integral over all defects having short relaxation times. For this reason, we have also included the values of $\delta v$ measured in the temperature region where the anharmonic contribution can be neglected, {\it i.e.} below 100~K. The results of the fits in the sonic and ultrasonic regimes are shown in Fig.~\ref{Fig4}. One finds $V_0\simeq$ 2460~K, $\tau_0\simeq$ 0.6 $\times 10^{-14}$~s and $\Delta_0\simeq$ 180~K. The two first values are in good agreement with those found in Ref.~\onlinecite{HERT1998}. The result for $\Delta_0$ cannot be compared with Ref.~\onlinecite{HERT1998} as the authors used a distribution with a constant density. Considering the differences in profiles between experiments taken at almost same frequencies, the agreement is satisfactory. It must be noted that the small contribution of TLS to $Q^{-1}$ was calculated from parameters extracted from the literature~\cite{RAU1995,CARI2008b}. This contribution is shown in Fig.~\ref{Fig4} as a thin black line and was added to the calculated TAR value to allow comparison with experiment.

From the parameters determined for the TAR mechanism, it is now possible to calculate the contribution of this process to $Q^{-1}$ at Brillouin scattering frequencies. The result is shown as a dashed line in Fig.~\ref{Fig5}. It is clearly much below the measured values. By subtracting $Q_{\rm TAR}^{-1}$ and $Q_{\rm TLS}^{-1}$ from the measurement, one gets the anharmonic contribution, $Q_{\rm ANH}^{-1}$, shown by circles in Fig.~\ref{Fig5}. One finds that $Q_{\rm ANH}^{-1}$ increases with increasing $T$ up to about 300~K and is nearly constant above up to $T_{\rm g}$.

\begin{figure}
\includegraphics[width=8.5cm]{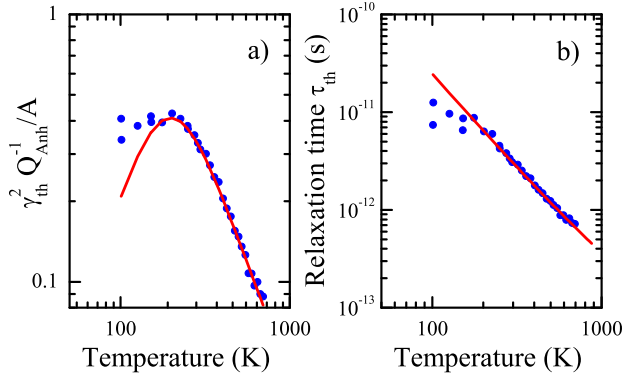}
\caption{(a) $Q_{\rm TAR}^{-1}$ scaled by $A(T)/\gamma_{\rm th}^2$. The solid line is obtained with Eq.~(\ref{Eq6}) using the fitted values of $\tau_{\rm th}(T)$ in~(b). (b) Temperature dependence of the thermal time $\tau_{\rm th}$. The solid line is a $T^{-2}$ power law adjusted to the data.}
\label{Fig6}
\end{figure}

The anharmonic contribution in crystals was calculated by \citet{MARI1971}. To analyze $Q_{\rm ANH}^{-1}$ in glasses, we follow a procedure described and discussed in detail in Ref.~\onlinecite{VACH2005}. $Q_{\rm ANH}^{-1}$ is given by:
\begin{equation}
Q_{\rm ANH}^{-1} = A \frac{\Omega \tau_{\rm th}}{1 + \Omega^2 \tau_{\rm th}^2},
\label{Eq6}
\end{equation}
with
\begin{equation}
A = \gamma_{\rm th}^2 \frac{C_v T v}{2\rho v_{\rm D}^3}.
\label{Eq7}
\end{equation}
Here, $C_v$ is the specific heat per unit volume and $v_{\rm D}$ is the Debye velocity, whereas $\tau_{\rm th}$ and $\gamma_{\rm th}^2$ are the average lifetime and the mean-squared average Gr\"uneisen parameter of thermal phonons, respectively. Except for the value of $\gamma_{\rm th}^2$, all the quantities in $A$ are known for $v$-GeO$_2$ according to Eq.~(\ref{Eq7}). The low-$T$ values of $C_v$ were taken from Ref. \onlinecite{ZELL1971}, complemented by the high-$T$ values of Ref. \onlinecite{DING1993}. We used $v_{\rm D}\simeq$ 2800 m/s and $\rho\simeq$ 3.65 g/cm$^3$. In Fig.~\ref{Fig6}a, we have plotted $\gamma_{\rm th}^2 Q_{\rm ANH}^{-1}/A$ as a function of $T$. From Eq.~(\ref{Eq6}), one expects a maximum of this quantity at a value $\gamma_{\rm th}^2/2$, when $\Omega\tau_{\rm th}$ = 1. From the points in Fig.~\ref{Fig6}a, we observe that $\gamma_{\rm th}^2 Q_{\rm ANH}^{-1}/A$ increases when $T$ decreases up to a shallow maximum around 0.4. From this value, we extract an evaluation of $\gamma_{\rm th}^2 \simeq$ 0.8. It is now possible to calculate $\tau_{\rm th}$ as a function of $T$. The result is shown in Fig.~\ref{Fig6}b. Except for the values below 200~K for which the uncertainty is large, the results agree with a variation $\tau_{\rm th} \propto T^{-2}$. The $T$ dependence of $\gamma_{\rm th}^2 Q_{\rm ANH}^{-1}/A$ can now be calculated: the result is shown  as a solid line in Fig.~\ref{Fig6}a. With the same parameters, it is possible to calculate $Q_{\rm ANH}^{-1}$ from Eq.~(\ref{Eq6}). The dotted line in Fig.~\ref{Fig5} shows the result, which is in good agreement with the estimation of this contribution extracted from the experiment. Finally, $Q_{\rm TLS}^{-1}$, $Q_{\rm TAR}^{-1}$ and $Q_{\rm ANH}^{-1}$ are added: the solid line in Fig.~\ref{Fig5} is obtained, in excellent agreement with the experiment.

\begin{figure}
\includegraphics[width=8.5cm]{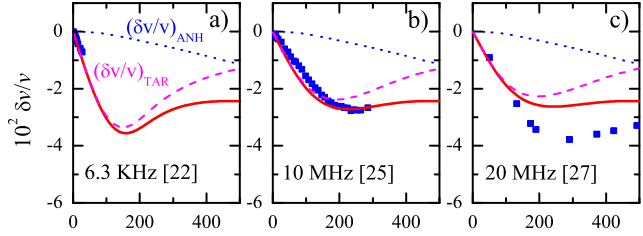}
\caption{(squares) Velocity changes of $v$-GeO$_2$ in the sonic and ultrasonic frequency ranges. The solid lines are the sum of TAR (dashed) and anharmonic (dotted) contributions as discussed in the text.}
\label{Fig7}
\end{figure}

From the known parameters for TAR and anharmonicity, it is possible to calculate the velocity changes produced by these mechanisms, $\delta v_{\rm TAR}$ and $\delta v_{\rm ANH}$, respectively (see Eq.~(\ref{Eq5}) for TAR and Eq.~(16) of Ref.~\onlinecite{VACH2005} for anharmonicity). The results are shown in Fig.~\ref{Fig7} where they are compared to the experimental values at sonic and ultrasonic frequencies. It must be noted that $\delta v_{\rm ANH}$ is nearly constant below 100~K, and that $\delta v_{\rm TAR}$ shows a deep minimum around 200~K, while $\delta v_{\rm ANH}$ decreases linearly in this region. On the three figures~\ref{Fig7}a-c, we remark that the slope of $\delta v(T)$ below the minimum is well reproduced. This slope is directly related to $\Delta_0$, the upper cut-off in the distribution of asymmetries. This indicates that only relaxing entities with asymmetry below about 180~K are present in $v$-GeO$_2$. On the other hand, the shallow minimum observed in the measurement at 10~MHz (Fig.~\ref{Fig7}b) is well reproduced. In contrast, the calculation disagrees with the experiment at 20~MHz. It must be noted that only a very small change of $\delta v$ is expected when the frequency varies from 10~MHz to 20~MHz as shown by the solid lines in Fig.~\ref{Fig7}b-c. Thus, we suggest that the disagreement at 20~MHz originates from uncertainties in the measurement.

\begin{figure}
\includegraphics[width=8.5cm]{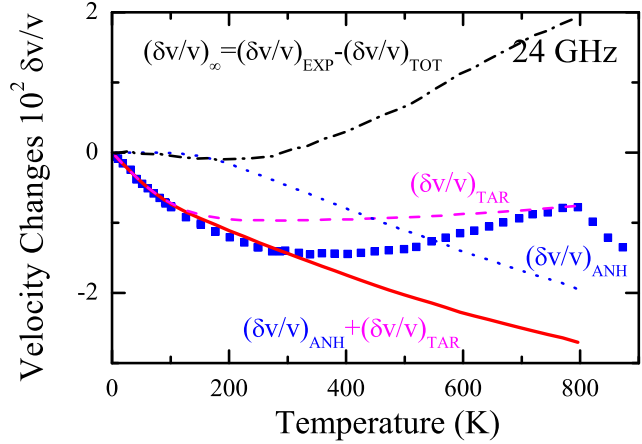}
\caption{(squares) Velocity changes in $v$-GeO$_2$ at Brillouin scattering frequencies. The dashed (dotted) line shows the calculated TAR (anharmonic) contribution whereas the solid line is the sum of these two mechanisms. The dashed-dotted line represents the relative changes of the calculated bare velocity $v_\infty$.}
\label{Fig8}
\end{figure}

Fig.~\ref{Fig8} shows the experimental and calculated values for $\delta v/v$ at Brillouin scattering frequencies, around 24~GHz. The velocity minimum observed here is much less deep by a factor of $\simeq$~2 than at ultrasonic frequencies. This is well described by the calculation, which is in excellent agreement with the experiment from the lowest $T$ to $\simeq$ 300~K. From this observation, and from the above remarks on the velocity slope, we conclude that the density of relaxing defects used for the fit is satisfactory. Finally, subtracting $\delta v_{\rm TAR}$ and $\delta v_{\rm ANH}$ from the experimental values, the bare velocity $v_\infty$ is obtained. In principle, one expects that $v_\infty$ is independent from $T$. We have plotted $(\delta v/v)_\infty$ versus $T$ as a dashed-dotted line in Fig.~\ref{Fig8}. While this quantity is constant from 0 to $\simeq$ 250~K, it shows a nearly linear increase above, up to $T_{\rm g}$. A similar behavior was already observed in $v$-SiO$_2$~\cite{VACH2005}.

To summarize, we have shown that a satisfactory description of velocity and attenuation measurements in a large $T$, $\Omega$ domain can be obtain by taking into account both TAR and anharmonicity mechanisms. We have shown that a rather low cut-off in the asymmetry distribution of TAR is needed for a good description of the velocity slope at low $T$. While the effect of anharmonicity on $Q^{-1}$ is negligible at ultrasonic frequencies, the contributions of both processes at Brillouin scattering frequencies have almost the same amplitude. This might explain why the velocity measurements in Ref.~\onlinecite{HERT1998} and the $Q^{-1}$ values at higher frequencies measured in UV Brillouin scattering in Ref.~\onlinecite{BALD2007} could not be described satisfactorily on the basis of the TAR model alone.

\section{Conclusions}

We have shown that the description of acoustic velocity and attenuation already established in $v$-SiO$_2$ is also appropriate for $v$-GeO$_2$. In both glasses, the effect of anharmonicity on the attenuation is negligible in the sonic and ultrasonic frequency range. The changes with frequency of the peak in $Q^{-1}$ observed in this region gives a value $V_0\simeq$ 2460~K for the activation energy in $v$-GeO$_2$, while $V_0\simeq$ 660~K was found in $v$-SiO$_2$, in agreement with the fact that the free-volume is smaller in $v$-GeO$_2$ than in $v$-SiO$_2$. The anharmonic contribution to $Q^{-1}$ becomes similar to the effect of TAR in the hypersonic regime. The effect of anharmonicity is negligible on the variation with $T$ of the sound velocity below 100~K. The latter can be described by assuming a cut-off in the asymmetry of TAR defects. This cut-off was found equal to $\simeq$ 90~K in $v$-SiO$_2$, while we find $\Delta_0\simeq$ 180~K in $v$-GeO$_2$. 

In $v$-SiO$_2$, it was shown that, when the velocity changes induced by TAR and anharmonicity are taken into account, the bare velocity is constant at low $T$, then increases linearly above $\simeq$ 150~K. When this model is applied to $v$-GeO$_2$, we find the same behavior: $v_\infty$ is constant up to 200~K and increases linearly above 300~K. In $v$-SiO$_2$, this anomalous temperature hardening was assigned to changes in the ring conformation induced by flips of Si$-$O$-$Si bonds~\cite{HUAN2004}. Our results indicate that a similar mechanism might be present in $v$-GeO$_2$. This could be an important matter to be studied by atomic simulations.

%

\end{document}